# A Comparison of Physical Properties and Fuel Cell Performance of Nafion and Zirconium Phosphate/Nafion Composite Membranes

Chris Yang[1,2], S. Srinivasan[3], A.B. Bocarsly[3], S. Tulyani[4] and J.B. Benziger[4,5]

Princeton University

Princeton, NJ 08544

The physio-chemical properties of Nafion 115 and a composite Nafion 115/Zirconium Phosphate (25wt%) membranes are compared. The composite membrane takes up more water than Nafion at the same water activity. However, the proton conductivity of the composite membrane is slightly less than that for Nafion 115. Small angle X-ray scattering shows the hydrophilic phase domains in the composite membrane are spaced further apart than in Nafion 115, and the composite membrane shows less restructuring with water uptake. Despite the lower proton conductivity of the composite membranes they display better fuel cell performance than Nafion 115 when the fuel cell is operated under-humidified. It is suggested that the composite membrane has a greater rigidity that accounts for its improved fuel cell performance.

1 Department of Mechanical and Aerospace Engineering

2 Current address University of California Davis

3 Chemistry Department

4 Department of Chemical Engineering

5 To whom inquires should be addressed

# INTRODUCTION

Polymer electrolyte fuel cells based upon perfluorinated membranes have typically been operated in a temperature range between approximately 50°C and 90°C.(Blomen and Mugerwa 1993; Srinivasan, Dave et al. 1993; EG&G Services 2000) This temperature range is a compromise between competing factors. Increasing the operating temperature above room temperature will improve the electrode kinetics of the oxygen reduction reaction.(Mukerjee and Srinivasan 1993) The upper limit of temperature results from the difficulty in maintaining membrane water content at temperatures at or above 100°C. In addition, temperatures above the polymer glass transition temperature (~110°C for protonated Nafion) can cause polymer chain rearrangements, which can lead to structural changes in the membrane and lower the membrane stability, performance, and lifetime (Yeo and Eisenberg 1977; Hinatsu, Mizuhata et al. 1994; Zook and Leddy 1996).

There has been a push to develop polymer membranes able to operate above 120°C prompted by the additional benefits of enhanced carbon monoxide (CO) tolerance and improved heat removal.(Yang, Costamagna et al. 2001)

The most significant barrier to running a polymer electrolyte fuel cell at elevated temperatures is maintaining the proton conductivity of the membrane. The increased temperature raises the evaporation rate of water from the membrane and the vapor pressure required to keep a given amount of water in the membrane; thereby increasing the likelihood that water loss will occur and significantly reduce proton conductivity. The conductivity of a dry membrane is several orders of magnitude lower than a fully saturated membrane. A number of alternative strategies have been investigated to maintain membrane conductivity in a dehydrating environment (i.e. elevated temperature and reduced relative humidity).

The addition of an inorganic material into a polymer membrane can alter and improve physical and chemical polymer properties of interest (such as elastic modulus, proton conductivity, solvent permeation rate, tensile strength, hydrophilicity, and glass transition temperature) while retaining its important polymer properties to enable operation in the fuel cell. A number of investigators have examined composite membranes for use in polymer electrolyte fuel cells (Malhotra and Datta 1997; Mauritz 1998; Antonucci, Arico et al. 1999; Grot and Rajendran 1999; Grot and Rajendran 1999; Honma, S.Hirakawa et al. 1999; Alberti, Casciola et al. 2000; Doyle, Choi et al. 2000; Lee, Adjemian et al. 2000; Mauritz and Payne 2000; Murphy and Cisar 2000; Miyake, Wainright et al. 2001; Adjemian, Lee et al. 2002; Honma, Nishikawa et al. 2002; Si, Lin et al. 2002). The hydration properties of membranes are key characteristics that can influence fuel cell performance. The composite membranes may improve the water-retention properties of these membranes under low humidity conditions. We have examined composite polymer/inorganic membranes (Nafion/Zirconium phosphate) in fuel cells at elevated temperatures. We have also characterized the physical, chemical and electrical properties of these membranes to see how fuel cell performance correlates with physical properties.

Zirconium hydrogen phosphate ($Zr(HPO_4)_2$) is a promising additive to Nafion based membranes because of its attributes, including: (i) it has moderate proton conductivity when humidified (~$10^{-3}$ S/cm) (Alberti 1978; Alberti, Casciola et al. 1984;

Casciola and Constantino 1986; Alberti, Casciola et al. 1992; Alberti, Casciola et al. 1992; Casciola, Marmottini et al. 1993; Alberti, Boccali et al. 1996; Alberti and Casciola 1997; Glipa, Lelout et al. 1997; Alberti, Casciola et al. 1999); (ii) it is a Bronsted acid with the ability to donate protons; (iii) it is thermally stability to temperatures above 180°C; (iv) it is hygroscopic and hydrophilic and (v) it is easily synthesized in a manner that is compatible with the chemical and physical limits of the polymer membrane.(Grot and Rajendran 1999)

The conductivity of perfluorinated sulfonic acid membranes vary over many orders of magnitude depending upon important parameters such as the water activity and temperature. Models for proton conduction in Nafion have been proposed that provide a good semi-quantitative prediction of the conductivity at water activities greater than 0.2.(Eikerling, Kharkats et al. 1998; Thampan, Malhotra et al. 2000)  However, very little has been done to characterize the conductivity of Nafion above 100°C or to characterize the proton conductivity in composite membranes (Miyake, Wainright et al. 2001).  We have compared water uptake, proton conductivity, microstructure morphology and fuel cell performance of Nafion and Zirconium Hydrogen Phosphate/Nafion composite membranes as functions of temperature (80-140°C) and water activity.

# EXPERIMENTAL

## Membrane Preparation

The composite membranes were prepared using Nafion® 115 films (DuPont) as the base material.  To obtain a uniform high purity film, membranes were cleaned with a standard treatment procedure: (i) boiling in 3% hydrogen peroxide for 1 hr to oxidize organic impurities; (ii) rinsing with boiling water for several hours;  (iii) boiling in 1 M sulfuric acid for 1 hr to remove any metallic/ionic impurities; and (iv) rinsing again in boiling water to remove any excess acid.

Zirconium phosphate was incorporated into Nafion using the procedure first described by Grot and Rajendran (Grot and Rajendran 1999).  The synthesis was of zirconium phosphate involves the reaction of a solution of $Zr^{4+}$ ions with phosphoric acid ($H_3PO_4$) leading to the precipitation of the insoluble zirconium phosphate.  In solution the procedure produces zirconium hydrogen phosphate, $(Zr(HPO_4)_2)$.  To accomplish this synthesis within the membrane, the procedure takes advantage of the fact that Nafion and other perfluorosulfonic acid ionomers are ion-exchange membranes.  The protons are exchanged with zirconium ions and the zirconium ions are subsequently reacted in place with phosphoric acid.

First, the membranes were weighed in the dry state, and then swollen in a boiling methanol-water solution (1:1 vol) to expand the membrane and facilitate ionic diffusion. The membranes were then dipped in a 1M solution of zirconyl chloride, $ZrOCl_2$ (Aldrich) for several hours at 80°C.  The $ZrOCl_2$ solution diffuses into the membrane and the excess of $Zr^{4+}$ ions within the membrane leads to an exchange with sulfonic acid protons in the membrane.  The membranes were then rinsed in cold water to remove the zirconyl chloride solution from the surface and then immersed in 1 M phosphoric acid ($H_3PO_4$) overnight at 80°C.  The phosphoric acid has two purposes: (i) it reacts with the $Zr^{4+}$ ions to precipitate insoluble zirconium hydrogen phosphate in-situ, and (ii) the acidic solution can re-protonate the sulfonate anions to regenerate the acidity of the membrane. The membranes were then repeatedly boiled for several hours in distilled water to remove any excess acid and $ZrOCl_2$ solution. After drying, membrane weight and thickness increased about 23% and 30% respectively as compared with the unmodified membrane. Other zirconium phosphate concentrations were also obtained (10-12% 20-25% and 36%) by varying the zirconyl chloride solution concentration.

## *Physical/Chemical Characterization*

The composite zirconium phosphate membranes were characterized an electron microprobe elemental analysis carried out using a Cameca SX50 experimental microprobe. The cross-sections of several Nafion and composite Nafion/zirconium phosphate membranes were analyzed using the electron microprobe to determine the presence and distribution of zirconium phosphate by measuring the distribution of phosphorus and zirconium. Micrographs revealed that the control Nafion contains negligible traces of these elements, while the composite membranes have a uniformly random distribution of zirconium and phosphorus throughout the cross-section.

Powder X-ray diffraction (XRD) patterns for these membranes were obtained at CNR-ITAE using a Philips X-Pert 3710 X-ray Diffractometer using Cu-$K_\alpha$-source operating at 40 kV and 30 mA.

Ion exchange capacity, IEC, was determined by an exchange of acidic protons with another cation in solution (Clearfield and Stynes 1964; Chen and Leddy 2000). The membranes were dried and weighed and then placed in a 1 M NaCl solution at 80°C overnight to exchange $Na^+$ ions with $H^+$. The large excess of $Na^+$ ions ensured virtually complete exchange. The membranes were removed from solution and the solution was titrated to the phenolphthalein end point with 0.1 M NaOH solution to determine the quantity of exchanged $H^+$ ions. The ion exchange capacity and equivalent weight (grams of polymer per mole of H+) were calculated using the dry weight of the polymer and the quantity of exchanged protons.

## *Water Uptake and Membrane Conductivity*

The water uptake measurements and the conductivity measurements, were carried out in a barometric sorption vessel that controlled temperature and humidification level. The barometric sorption apparatus, shown in Figure 1, is based on a design described by Miyake et al. (Miyake, Wainright et al. 2001).

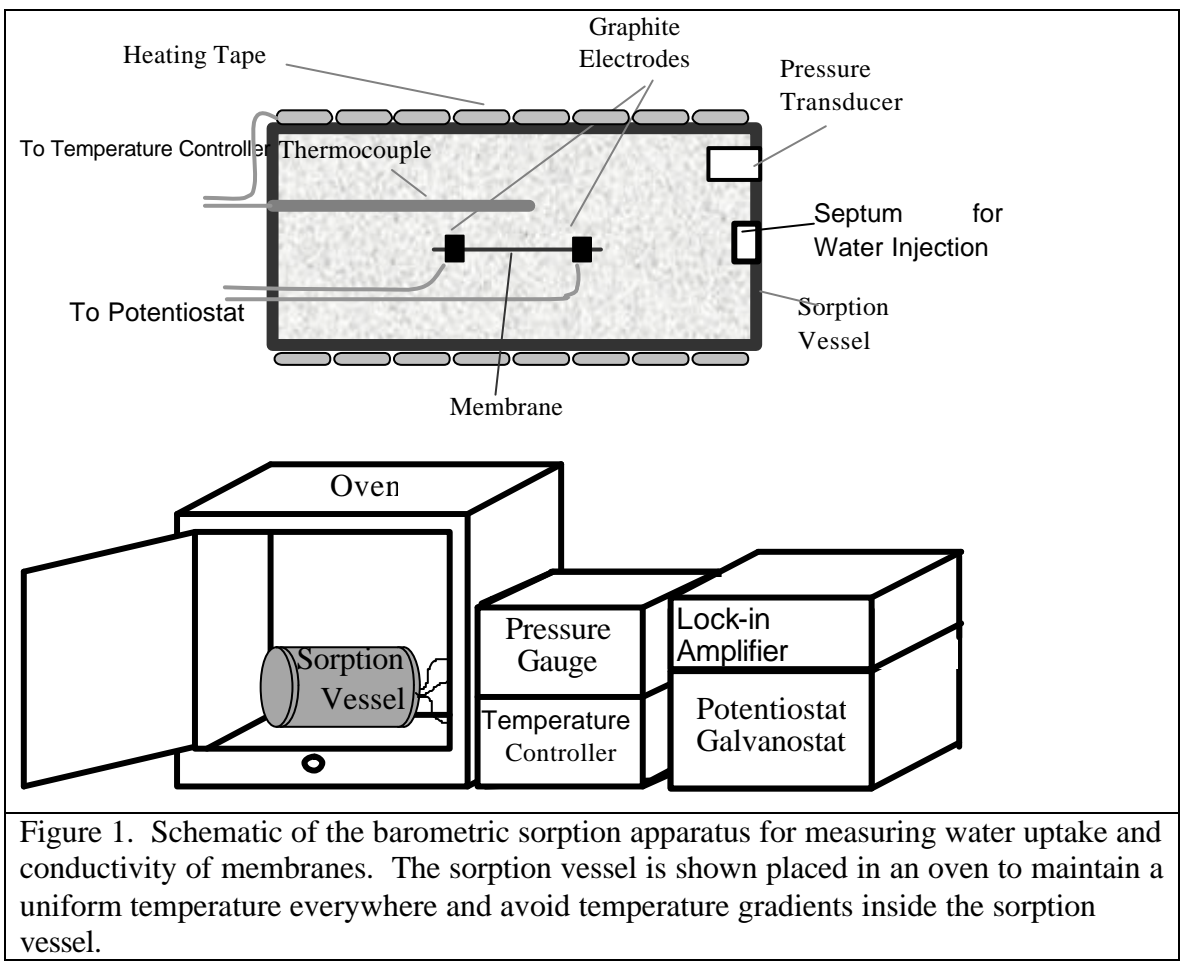

Figure 1. Schematic of the barometric sorption apparatus for measuring water uptake and conductivity of membranes. The sorption vessel is shown placed in an oven to maintain a uniform temperature everywhere and avoid temperature gradients inside the sorption vessel.

The sorption vessel volume is 430 mL and is equipped with an Omega pressure transducer, copper and thermocouple wire feedthroughs, and swagelok® fittings with a septum for water injection. The unconstrained dry membrane is placed within the sorption vessel, which is evacuated at the measurement temperature (80-140°C) for over an hour to dry the membrane and remove any residual water from the vessel. Water is injected into the vessel through the septum using a microliter syringe.

At temperatures between 80 and 140°C, the small quantity of water that is injected (typically 5-50 μl) evaporates quickly and increases the pressure in the vessel. Using the ideal gas law, the expected pressure, $P_{exp}$, associated with the vaporization of the injected water can be calculated.

$$P_{\exp} = \frac{m_{H2O}RT}{18V} \tag{1}$$

where $m_{H2O}$ is the mass of injected water and V is the vessel volume.

$$N_{H2O} = \frac{V(P_{\exp} - P_{act})}{RT} \tag{2}$$

The difference between the expected pressure and the actual measured pressure, $P_{act}$, is attributed to a condensed phase of water in the membrane and the number of moles of water in the membrane, $N_{H2O}$, can be calculated. The number of moles of sulfonic acid in the membrane is calculated from the dry weight of the membrane, $m_{mem}$, and equivalent weight, $EW$.

$$N_{SO3^-} = \frac{m_{mem}}{EW} \tag{3}$$

Finally the membrane water content parameter, $\lambda$, the number of moles of water per mole of sulfonate can be calculated.

$$\lambda = \frac{N_{H2O}}{N_{SO3^-}} \tag{4}$$

After allowing at least 30 minutes for membrane equilibration with the vapor phase, another small quantity of water is injected and the pressure is measured again. This procedure is repeated until no pressure rise is detected with the water injection, which indicates the saturation vapor pressure, $P_{sat}$, is reached.

Within the sorption vessel, the membrane conductivity was measured using an AC impedance technique to isolate the bulk membrane resistance from other resistance factors.

The conductivity of the perfluorinated sulfonic acid membranes is measured along the longitudinal direction (in plane) of the membrane. A small piece of the membrane is placed between a set of graphite blocking electrodes spaced 1 cm apart. The graphite electrodes are connected through the wire feedthroughs to measurement equipment. The two probe frequency dependent impedance measurement was carried out on a Princeton Applied Research (PAR) Model 398 Electrochemical Impedance System, consisting of a potentiostat/galvanostat Model 273A and a lock-in amplifier Model 5210 which are connected to a PC running Electrochemical Impedance Software (EIS). The applied signal is a single sine wave of 5 mV with frequencies varying between $10^5$ Hz to 10 Hz. Both water uptake and conductivity measurements are made by injecting small quantities of water into the sorption vessel and recording the quantity of water injected, the pressure in the sorption vessel and membrane resistance. The water activity was evaluated from the actual pressure in the sorption vessel divided by the saturation pressure.

## Small Angle X-ray Scattering

### SAXS Sample Preparation

Nafion membranes and nafion/zirconium phosphate composite membranes were used after the cleaning treatment procedure described above. The thickness of the sample was chosen to allow for approximately 40% transmittance. In some cases, it was necessary to use two pieces of membrane to obtain the optimal thickness. Before each set of runs, the membrane was placed in boiling de-ionized water to ensure full hydration. The membrane was then placed in a viton sealed copper sample cell with mica windows. The cell was opened and the membrane was slowly dehydrated between runs to allow for scattering data collection at different water contents. Membrane water content, ?, was determined gravimetrically.

**SAXS Data Collection and Analysis**

The sample cell was placed in the path of the x-ray beam. Membrane water loss during the data collection was minimized by filling the flight tube with helium rather than a vacuum. The 1.54 Å Cu-Kα x-rays were generated by a Philips XRG-3000 sealed tube generator source. The beam was slit collimated and the scattering was detected by an Anton-Paar compact Kratky camera equipped with a Braun OED-50M detector. Samples were typically run at room temperature for 10 minutes. Background beam scattering, sample transmittance, and detector response were corrected for in the data analysis. The data reduction and desmearing procedures are described in detail by Register.(Register and Bell 1992) The invariant scattering intensity ($q^2I$) is plotted as a function of scattering angle or distance and the Bragg spacing is determined by the location of the peak in scattering intensity.

## Water Transport

The water flux through different membranes at 80°C was measured as liquid water was fed to the one side of the membrane, and a dry gas was flowed to the opposite side. Nafion 115, Nafion 117 and Nafion 115/Zirconium phosphate membranes were prepared as previously described. Each of these membranes was hot-pressed at 135°C without electrodes, between sealing gaskets for 2 minutes at 10 kN (1 metric ton).

The membrane was placed in a 5 cm$^2$ fuel cell housing with triple-pass serpentine flow fields and heated to 80°C. Liquid water was passed through the flow field on one side of the membrane, and a dry $N_2$ stream was passed through the other side. The water vapor in the Nitrogen outlet was condensed in a cold trap at 0ºC and collected in a graduated cylinder. The rate of water flux through the membrane was calculated and plotted against the nitrogen flow rate.

## Fuel Cell Tests

**Membrane Electrode Assembly (MEA) Preparation**

Commercial gas-diffusion electrodes (20% Pt-on-carbon, 0.4 mg Pt/cm$^2$, purchased from E-TEK) were brushed with 5 wt% solubilized Nafion (Aldrich) to impregnate the active layer (0.6 mg/cm$^2$) and then dried at 80°C for 1 h. The geometrical area of the electrodes was 5 cm$^2$. A membrane was sandwiched between two electrodes and gas sealing gaskets, and the membrane-electrode assembly (MEA) was then pressed for 2 min at 135ºC at 20 MPa using a Carver hot press.

**Single Cell Test Fixture and Performance Evaluation**

The MEAs, coupled with gas-sealing gaskets, were placed in a single cell test fixture. The $H_2$ and $O_2$ (BOC) gases were fed to the single cell at 100 sccm. The gases were bubbled through water in temperature-controlled stainless steel bottles to humidify the feeds prior to entry to the fuel cell. The baseline test was total pressure of 1 bar, cell temperature of 80°C, and the humidifier bottles at $T_{anode} = 99°C$ and $T_{cathode} = 88°C$ respectively. Performance evaluations were carried out at 120-140ºC with a backpressure regulators at the effluents from the fuel cell fixed at 3 bar.  The temperatures of the humidifier bottles were varied to alter the water activity of the feed.

The fuel cell performance was characterized by current-voltage measurements (polarization curves).  These were recorded at 80°C and atmospheric pressure as well as in the range of temperatures between 80°C and 140°C, and total pressure of 3 bar pressure.  The maximum temperature of the humidifier bottles was 130°C, corresponding to a maximum water vapor pressure of 2.65 bar.  The fuel cell was preconditioned by operating at  0.3 V,  and high current density prior to the performance measurement.

Pseudo-steady state current-voltage measurements were obtained by connecting the fuel cell to a load resistance (either carbon film resistors or an electronic Amrel load), and allowing the current and voltage output of the single cell to settle to fixed values (~5-20 sec).  After the values of current and voltage were recorded, a new load condition was used and the single cell output was recorded.  The measurements were made starting at open circuit (zero current) and increasing current with each subsequent load condition.  Because the entire current-potential curve for a given temperature/humidification condition is obtained in a few minutes, it is assumed that the membranes have constant water content throughout the measurement.

# RESULTS

## Physical/Chemical Characterization

Table 1 compares the density change and ion exchange capacity for Nafion 115 and the nafion/zirconium phosphate composite membrane.  The density of the composite membrane is significantly lower than expected based upon the density of Nafion and zirconium phosphate (2.1 g/cm$^3$).  The low density of the composite membrane suggests it has void volume.  Some residual water associated with zirconium phosphate, which is not driven off except at very high temperatures, may also increase the volume of the composite membrane.

Table 1
Physical Characteristics of Nafion and Nafion/Zirconium Hydrogen Phosphate
Composite Membranes

| Membrane | Thickness ($\mu m$) | Density ($g/cm^3$) | IEC [$meq/g$] | EW [$g /mol H^+$] |
|---|---|---|---|---|
| Nafion 115 | 130 | 2.0 | 996 | 1004 |
| Nafion/zirconium phosphate (25%) | 170 | 1.6 | 1464 | 683 |

XRD patterns were obtained for the composite membrane in varying states of hydration: (i) a fully hydrated membrane (equilibrated with liquid water), (ii) a partially hydrated membrane and (iii) a thoroughly dried membrane.   The hydrated membrane shows only one weak peak circa 2.64 Å, whereas the well-dried membrane shows several sharp, well-defined peaks (Figure 2).  Diffraction maxima for the composite membrane may be attributed to Teflon-like domains of Nafion (5.2 Å and 2.3 Å) (Arico, Creti et al. 1998), and others that are attributed to the presence of zirconium phosphate phases (4.5, 3.73, 2.64, 1.7 Å).  Some of the zirconium phosphate peaks match up with typical peaks found in alpha zirconium hydrogen phosphate hemihydrate (4.46, 3.54 Å) (Alberti, Costantino et al. 1994).  The X-ray patterns are not very sharp, and water uptake by the membrane disrupts the crystalline microstructure. It is likely that the membrane swelling due to the uptake of water molecules influences the crystalline packing of polymer chains as well as the structure of the zirconium phases.

Diffraction peak width was used to estimate the particle size of the zirconium phosphate by using the Debye-Sherrer formula with Warren correction for instrumental effects.  In the fully dried membrane the zirconium phosphate particles are estimated to be 11 ±1 nm in size.

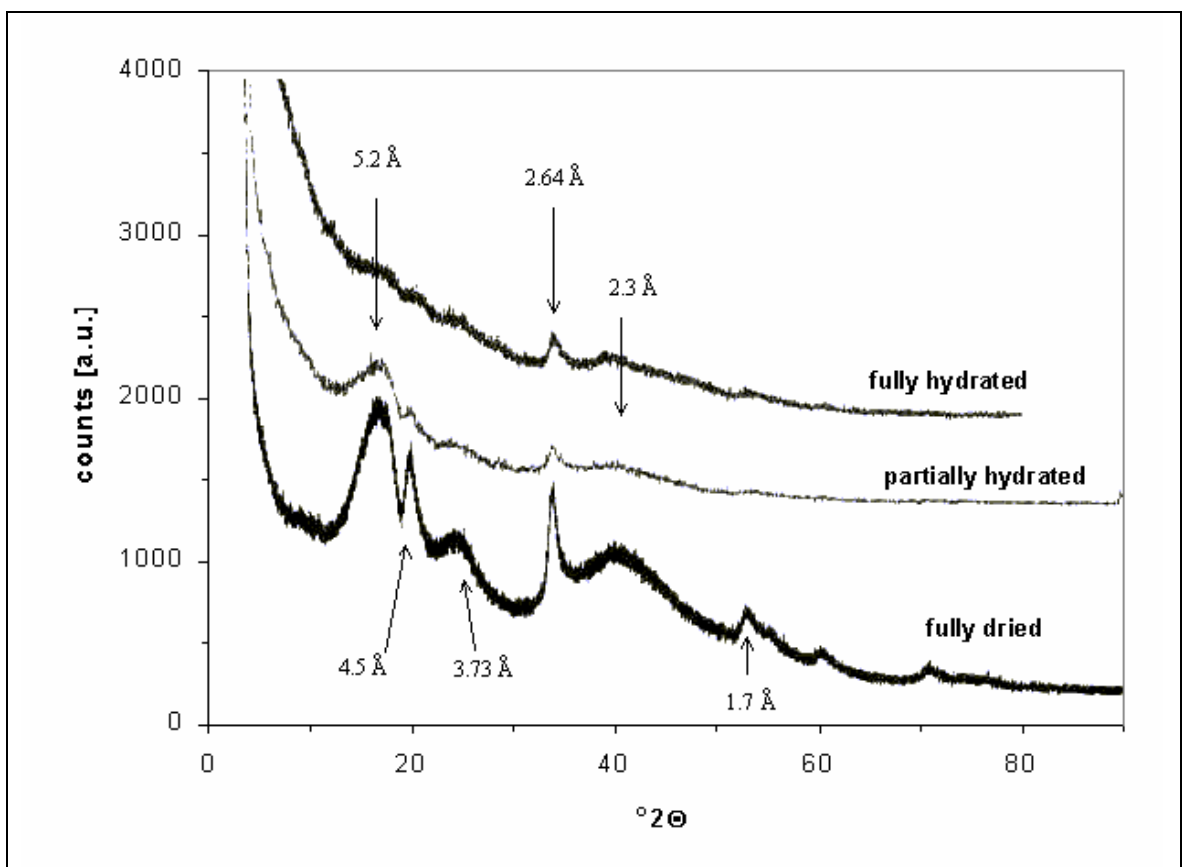

Figure 2.  X-ray diffraction patterns for Nafion 115/Zirconium phosphate (25 wt%) composite membranes.  The broad peaks at 2.3 and 5.2Å are attributed to crystalline stacking in the Teflon rich microphases.  The sharper features at 1.7, 2.64 and 4.5 Å are attributed to ziroconium hydrogen phosphate.

The ion exchange capacity and equivalent weight (grams of polymer per mole of H+) were calculated using the dry weight of the polymer and the quantity of exchanged protons.  Table 1 shows the results of ion exchange experiment.  The composite zirconium phosphate membranes have an increased ion exchange capacity (~40-50%) as compared to Nafion. Zirconium hydrogen phosphates have protons that can be exchanged which give the composite membrane a much greater ion exchange capacity.

## Water Uptake and Conductivity

Table 2 summarizes the water content of Nafion and Nafion/ZP composite membranes equilibrated with 100% relative humidity air at 80°C and immersed in liquid water at 25°C. Both these conditions correspond to water activity of 1. The membranes were removed from the water or the barometric sorption vessel and surface water is brushed off to get an accurate membrane water content measurement gravimetrically. The water content, $\lambda$, is calculated from the mass of water and equations 4 and 5. Even though 100% relative humidity and liquid water both correspond to water activity of one there is a difference in the water uptake. Liquid water uptake is greater than water vapor uptake. This difference has been attributed to the osmotic pressure of the membrane swelling in the liquid that can increase water content considerably (Schroeder's paradox).

Table 2
Saturation Water Uptake by Nafion and Nafion/ZP Composite Membranes

| membrane treatment | liquid water uptake (25°C) | | water vapor uptake (80°C) | |
|---|---|---|---|---|
| | weight % | $\lambda$ ($H_2O/SO_3H$) | weight % | $\lambda$ ($H_2O/SO_3H$) |
| Nafion 115 | 41 | 25 | 18 | 11 |
| Nafion 115 ZP (25%) | 33 | 25 | 25 | 19 |

The water uptake isotherms at 80°C for Nafion 115 and the Nafion/ZP composite membranes are shown in Figure 3. As the water activity in the vessel is increased, the membrane water content increases. The isotherm show a rapid rise at low water activity, a slow rising plateau at intermediate water activities and the majority of water uptake occurring at high water activity $a_w > 0.6$. At 80°C and water activity, $a_w = 1$, the Nafion membrane contains around 11 waters per sulfonate or about 18 wt. % of water. The solid line in Figure 3 shows the results of the BET finite layer isotherm model describing the membrane water content dependence on relative humidity.

While the uptake of water from the extruded Nafion membranes has been characterized in the literature (Hinatsu, Mizuhata et al. 1994; Futerko and Hsing 1999), water uptake by composite membranes has not received much attention. As seen in Figure 3 the composite membrane contains more water than a Nafion membrane at the same water activity. This number of waters is based only on the sulfonic acid content in the Nafion and neglects the interaction of water with the Zirconium phosphate. However, even on a total weight basis the Nafion/ZP membrane absorbs more water than Nafion.

The water uptake was not very sensitive to the temperature over the range of 80-140°C. Our data could not discern any significant difference in the water uptake with temperature for fixed water activity.

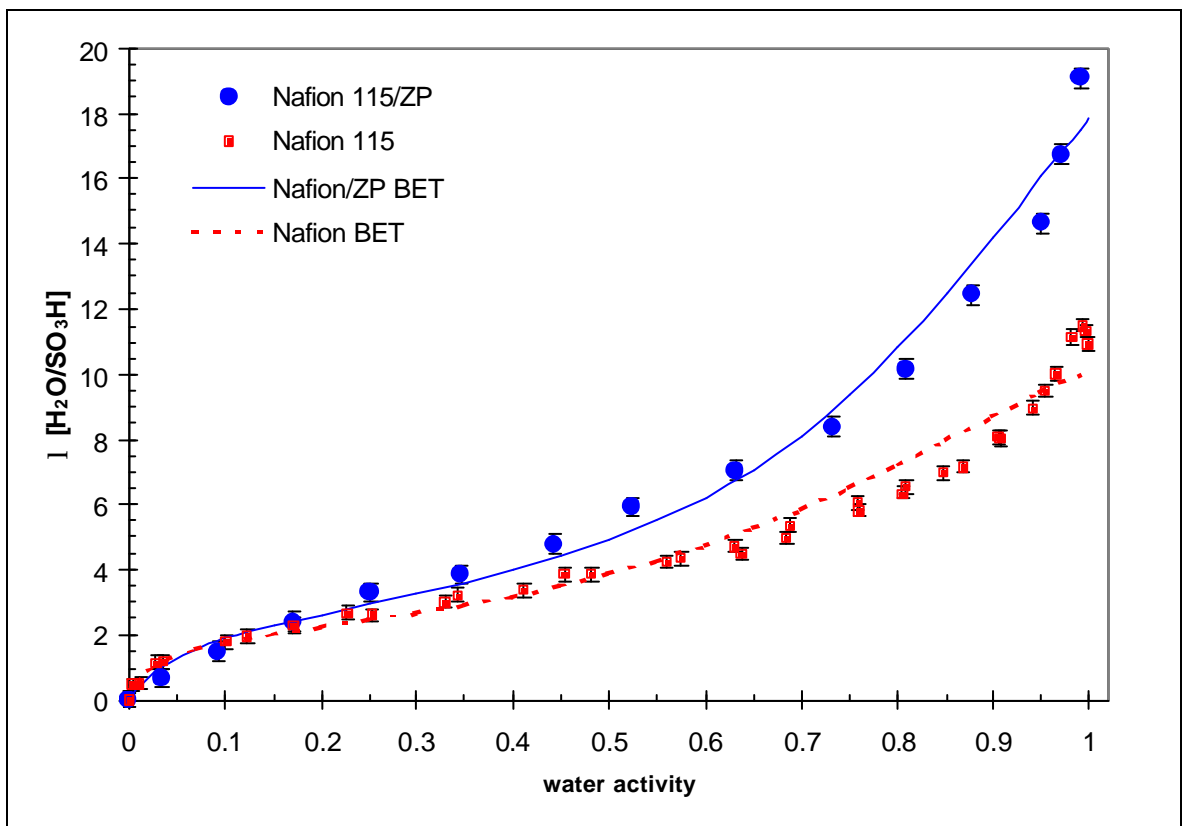

Figure 3. Water uptake of Nafion115 and Nafion115/ZP(25wt%) composite membranes at 80°C. The data were fit to finite layer BET isotherms (equation ) with fitting parameters: Nafion 115 ($\lambda_m$=2, c=35, $n_L$=9.2) Nafion/ZP composite ($\lambda_m$=2.6, c=17, $n_L$=12.8)

The conductivity of Nafion 115 as a function or water activity over a range of temperatures from 80°C to 140°C is shown in Figure 4. The data has been plotted on both log and linear scales. Most obvious from the log graph is the significant change in conductivity with water activity. The conductivity increases by five orders of magnitude. The conductivity values at $a_w$=0 are probably high because of incomplete water removal during the initial evacuation of the barometric sorption vessel can have a large impact. Based on the total pressure measurement the water activity is less than 0.002 for low conductivity values. The most reliable conductivity measurements are for intermediate water activity, 0.15<$a_w$<0.95 and the conductivity increases by more than two orders of magnitude with increasing water activity over that range. The temperature effect on conductivity is much smaller than the effect of water activity. By plotting the data on a linear scale the temperature effect becomes more evident. There is approximately a factor of two increase in conductivity from 80°C to 140°C at fixed water activity.

We expected Nafion/zirconium phosphate membranes to have an increased conductivity compared to unmodified Nafion. The water content was higher in the composite membrane and the zirconium phosphate contributed extra protons to increase the charge carrier concentration. Figure 5 compares the conductivity of the Nafion/Zirconium phosphate composite membrane with Nafion 115. The composite membranes had lower proton conductivity than Nafion over the entire range of water activities and temperatures.    Similar to Nafion 115, the major conductivity change for the composite membrane is associated with the water activity.  The conductivity of the composite membranes increased with temperature, but the variation of the composite membrane conductivity with temperature was slightly less than that observed with Nafion 115.

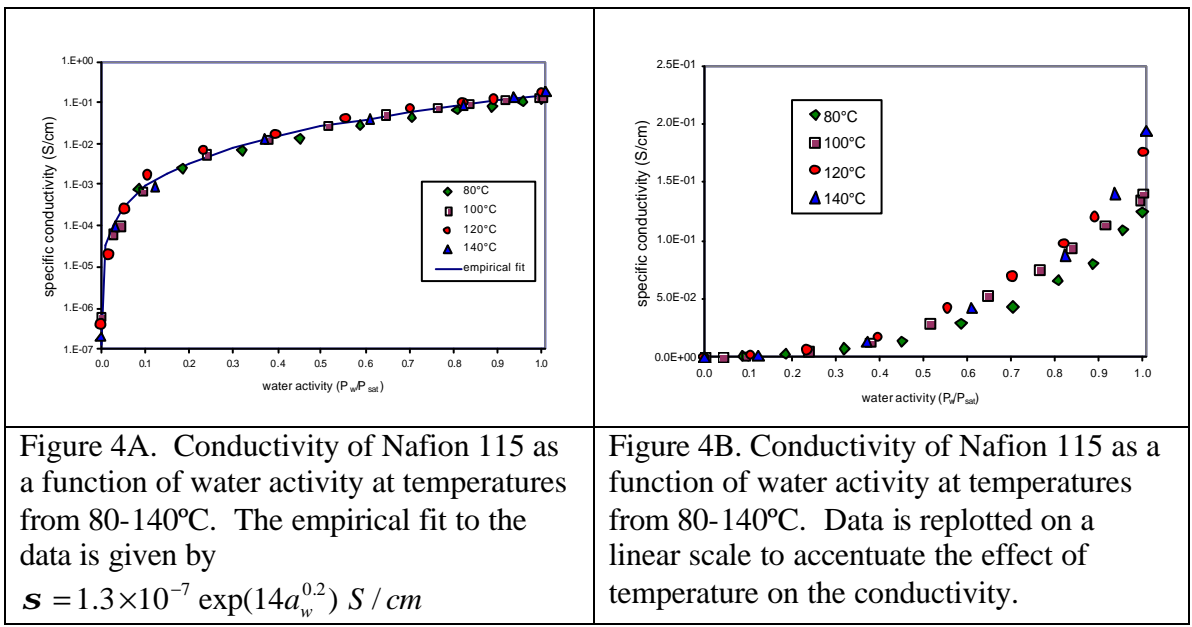

Figure 4A.  Conductivity of Nafion 115 as a function of water activity at temperatures from 80-140ºC.  The empirical fit to the data is given by

$$\boldsymbol{s} = 1.3 \times 10^{-7} \exp(14 a_w^{0.2}) \; S/cm$$

Figure 4B. Conductivity of Nafion 115 as a function of water activity at temperatures from 80-140ºC.  Data is replotted on a linear scale to accentuate the effect of temperature on the conductivity.

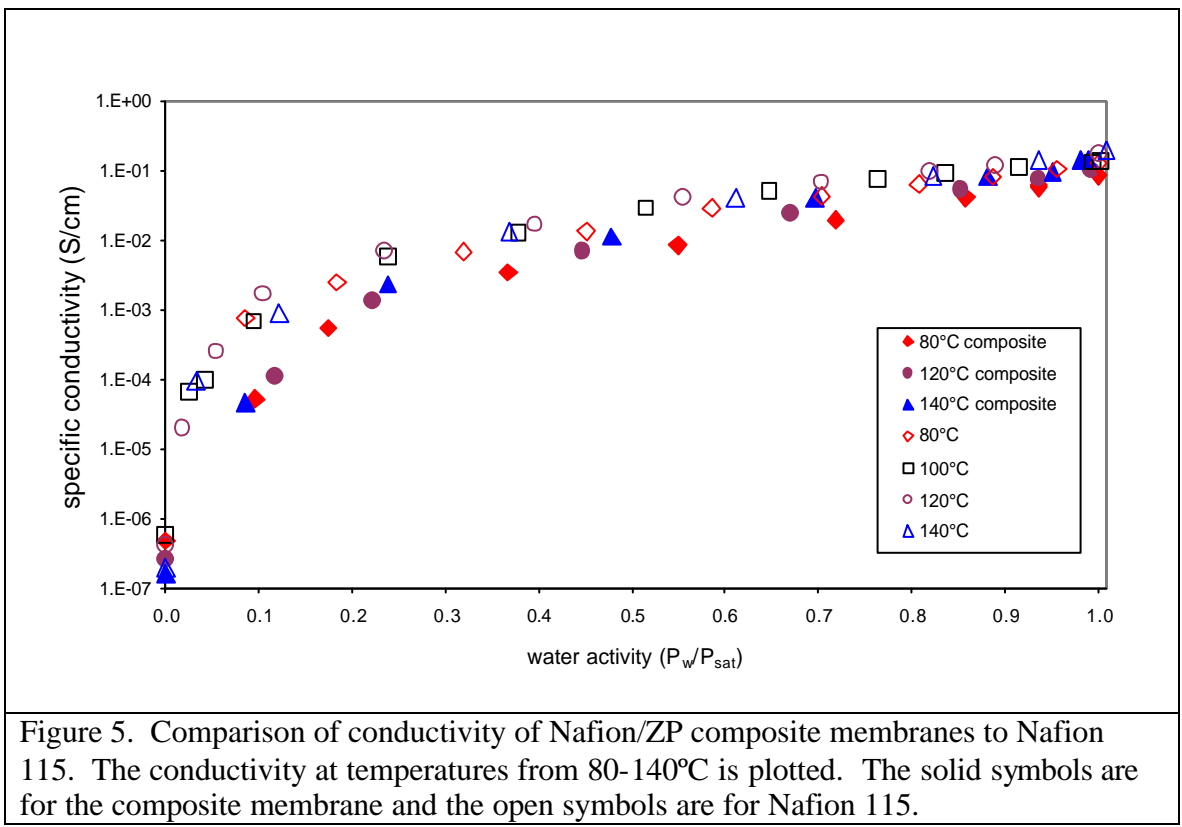

Figure 5.  Comparison of conductivity of Nafion/ZP composite membranes to Nafion 115.  The conductivity at temperatures from 80-140ºC is plotted.  The solid symbols are for the composite membrane and the open symbols are for Nafion 115.

## *SMALL ANGLE X-RAY SCATTERING*

The slit collimated scattering data collected by the multi-channel detector was converted into the scattering intensity vs. scattering angle.  The raw data was modified to account for several factors:

(i)       The channel detector that corresponds to zero angle was determined, and scattering due to interaction with the Helium sample environment was subtracted.

(ii)      The absolute scattering intensity of the sample was determined by comparing it quantitatively with a standard polyethylene sample.

(iii)     The sample was slit collimated in order to increase scattering intensity, by passing the beam over a larger sample area.  The data was desmeared using an algorithm from Lake.

(iv)     The invariant scattering intensity ($q^2 I$) was plotted as a function of scattering angle or distance.  The Bragg spacing is determined by the location of the peak in scattering intensity.

The invariant scattering intensity profile is plotted against scattering angle and the resultant peak intensity is correlated to an average Bragg spacing for each membrane sample as a function of water content, $\lambda$.  Ionomers, such as Nafion, will micro-phase separate into a Teflon-like region and an ionic region containing the sulfonic acid groups.(Yeo and Eisenberg 1977; Gierke, Munn et al. 1982; Hsu and Gierke 1982; Kreuer 2001)  Water partitions into the ionic regions.  Water has a significantly lower electron density than the fluorocarbon polymer matrix.  As a result, the electron density contrast between the ionic and Teflon-like phases of Nafion increases with increasing water content, resulting in greater scattering intensity and an increasingly well-defined scattering peak.  At low water contents ($\lambda < 4$ $H_2O/SO_3H$), the electron density contrast is small and a distinct scattering peak cannot be observed.(Roche, Pineri et al. 1981)

In figure 6, the invariant scattering intensity ($q^2 I$) is plotted against the Bragg spacing for an extruded Nafion 115 and a Nafion 115/ZP composite membrane. The data obtained shows that the as the increasing water content leads to an increase in the intensity of the scattering peaks.  The scattering intensity increases linearly with water content.  In the pure Nafion membranes, the electron density of the ionic inclusions is slightly lower than that of the polymer backbone while in the zirconium phosphate composite membrane the electron densities of the two phases are almost equal. For this reason there is little electron density contrast between the two phases when the membranes are dry and no scattering peak is obtained.  As the water content increases, the electron density of the ionic inclusions decreases and the intensity of the peaks increases.(Gierke, Munn et al. 1982)

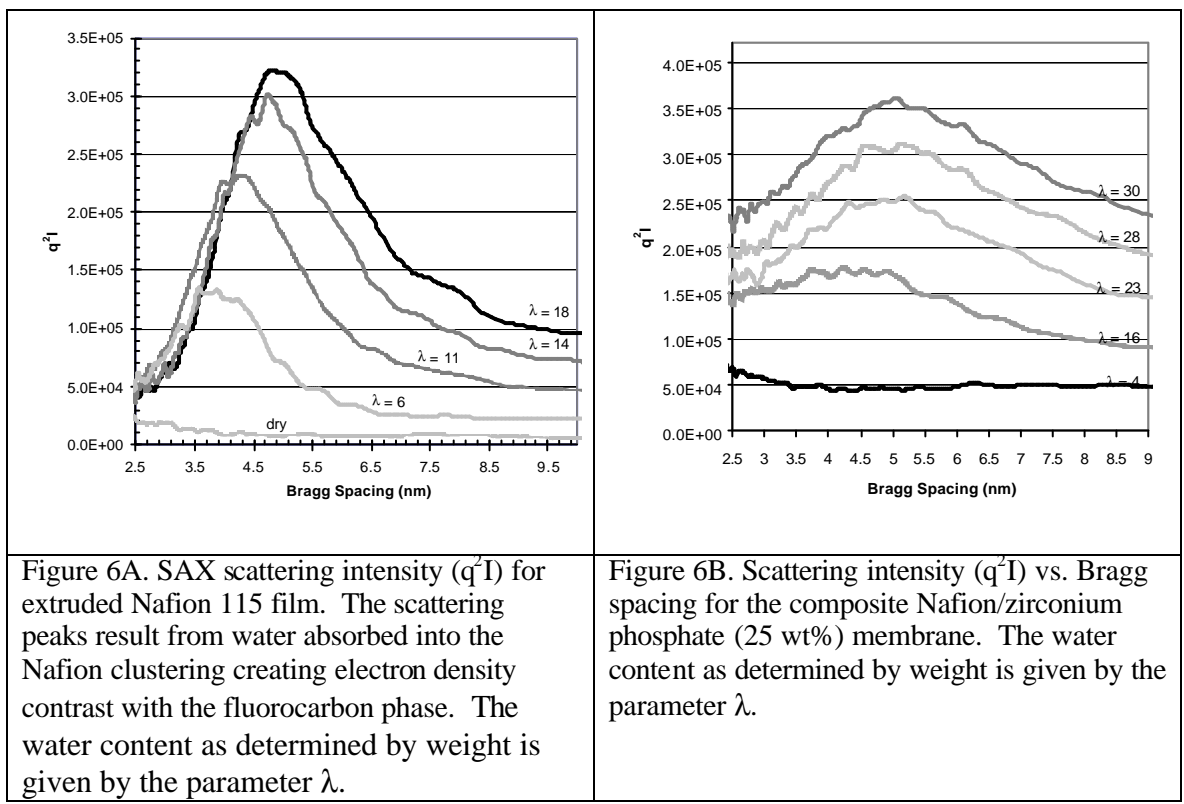

| Figure 6A. SAX scattering intensity ($q^2I$) for extruded Nafion 115 film. The scattering peaks result from water absorbed into the Nafion clustering creating electron density contrast with the fluorocarbon phase. The water content as determined by weight is given by the parameter $\lambda$. | Figure 6B. Scattering intensity ($q^2I$) vs. Bragg spacing for the composite Nafion/zirconium phosphate (25 wt%) membrane. The water content as determined by weight is given by the parameter $\lambda$. |
|---|---|

The scattering peaks for the zirconium phosphate membrane in figure 6B are significantly wider than the peaks for the unmodified Nafion 115 membrane, indicating greater structural heterogeneity. In addition, the maximum scattering intensities are reduced as compared to unmodified Nafion 115. The maximum spacing between ionic inclusions at the highest water content is nearly the same for the Nafion 115 and the Nafion 115/ZP composite. However, the shift in the spacing with water content is much less for the composite membrane than for Nafion 115.

## *Water Transport*

The flux of water through the membranes as a function of the nitrogen flow through the test cell is shown in Figure 7. The Nafion 117 membrane has almost the same thickness as the Nafion 115/ZP composite membrane, permitting us to distinguish between the effects of membrane thickness and membrane composition. The data in Figure 7 show the water flux increases with nitrogen flow rate through the test cell at low flow rates and then plateaus at a limiting flux. The limiting flux corresponds to the minimum mass transfer resistance at the membrane gas interface, and diffusion through the membrane is the dominant mass transfer resistance. The limiting flux is greatest for the Nafion 115 membrane. The flux was reduced through the Nafion 117 membrane due to increased membrane thickness. The limiting water flux through the composite membrane is less than that for Nafion 117 suggesting that water diffusion through the composite membrane is reduced relative to diffusion through Nafion.

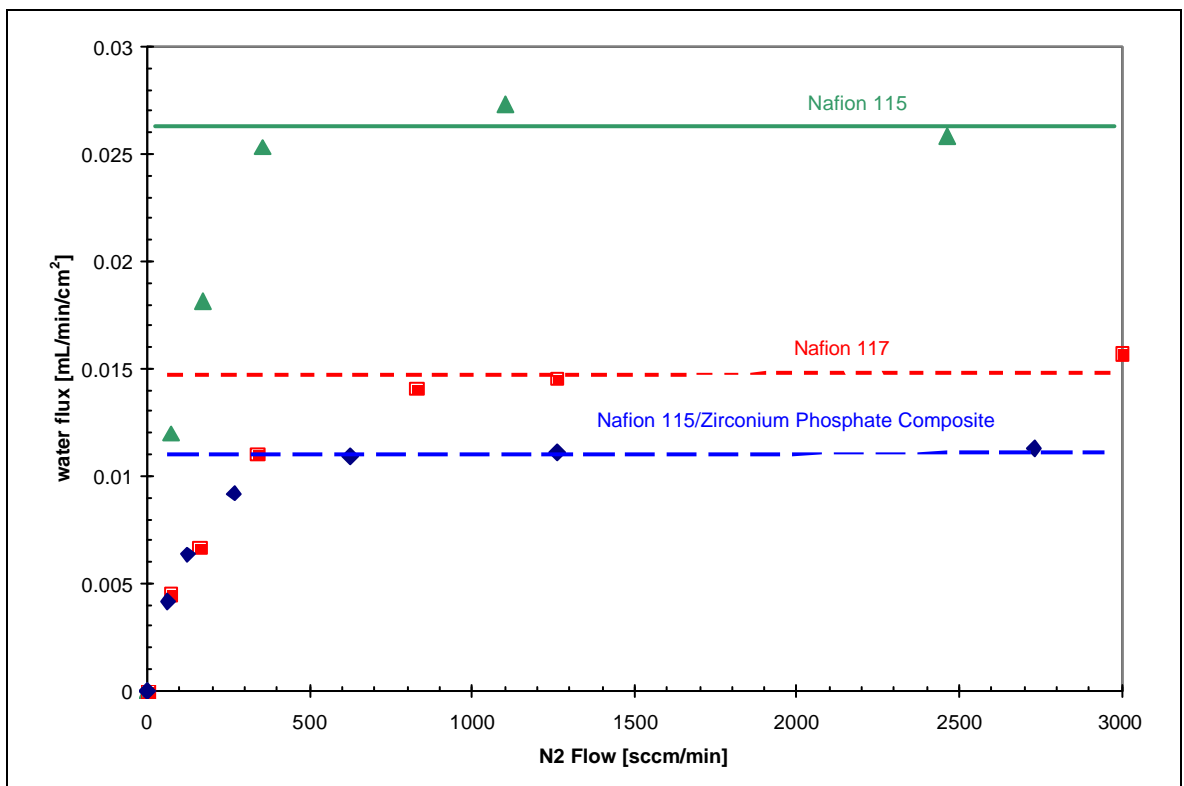

Figure 7.  Water flux through Nafion 115, Nafion 117 and Nafion 115/Zirconium Phosphate Composite membranes as functions of gas flow on the opposite side of the membrane.  The limiting fluxes are denoted by the horizontal lines.

## *Fuel Cell Performance of Nafion/Zirconium Phosphate Composite Membranes*

Polarization curves for fuel cells with membrane electrode assemblies (MEA) containing Nafion 115 and Nafion 115/Zirconium Phosphate membranes were measured at four different sets of operating conditions.  The base case was for humidified feeds near water activity of one with a cell temperature of 80°C.  The other three conditions were at elevated temperature and a total pressure of 3 bar.  The conditions tested were:

Operating Conditions:
    $1 - P=1$ bar, $T_{anode}/T_{cell}/T_{cathode}=90/80/88$
    $2 - P=3$ bar, $T_{anode}/T_{cell}/T_{cathode}=130/120/130$
    $3 - P=3$ bar, $T_{anode}/T_{cell}/T_{cathode}=130/130/130$
    $4 - P=3$ bar, $T_{anode}/T_{cell}/T_{cathode}=130/140/130$

Polarization curves are shown in Figure 8 for fuel cells with Nafion 115 and Nafion 115/Zirconium Phosphate membranes at conditions 1 and 3.  At 80°C the fuel cell performance was nearly identical for both membranes, with the effective MEA resistance being slightly greater for the composite membrane than Nafion 115.  When the temperature was 130°C the MEA resistance was substantially less for the composite membrane than for the Nafion 115 membrane.

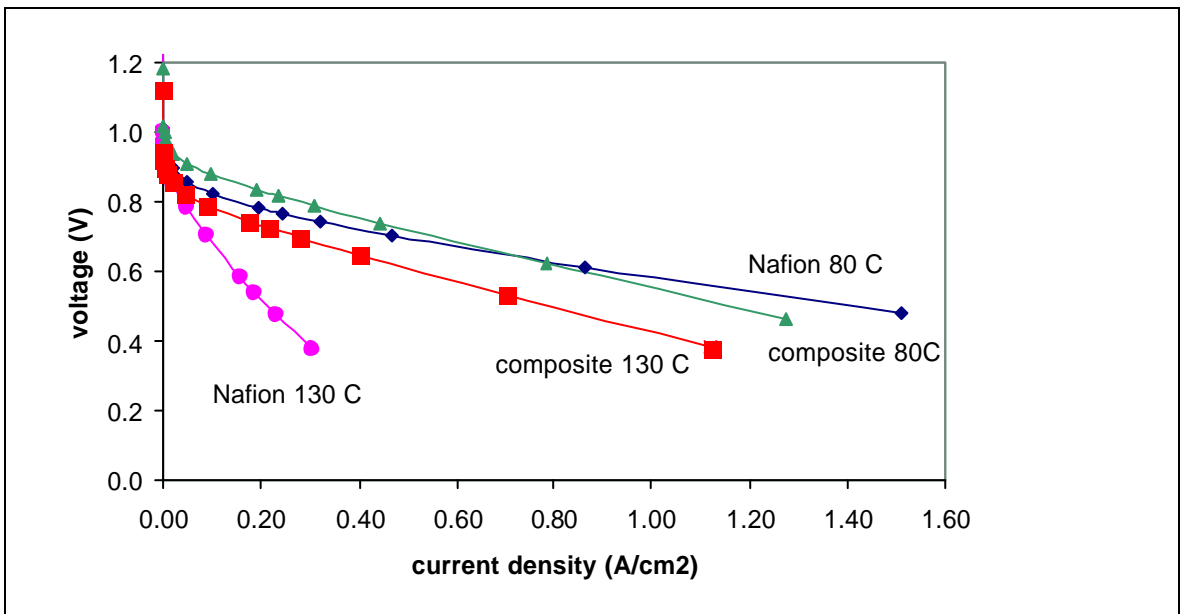

Figure 8.  Comparison of PEM fuel cell performance with MEA containing Nafion 115 and Nafion 115/Zirconium Phosphate composite membranes.

In Figure 9 we show the polarization curves for fuel cells with Nafion 115 and Nafion 115/Zirconium Phosphate Composite membranes at different levels of humidification.  The temperature designation $T_{anode}/T_{cell}/T_{cathode}$ corresponds to the temperatures of the humidification bottles for the anode and cathode and the fuel cell temperature.  By maintaining the humidification temperatures fixed the water vapor pressure of the feeds are fixed.  As the cell temperature is increased the water activity in the cell is decreased, $a_w = P_w/P_{sat}(T_{cell})$.  The effective MEA resistance increased less with decreased water activity for the composite membrane than observed with the Nafion 115 membrane.

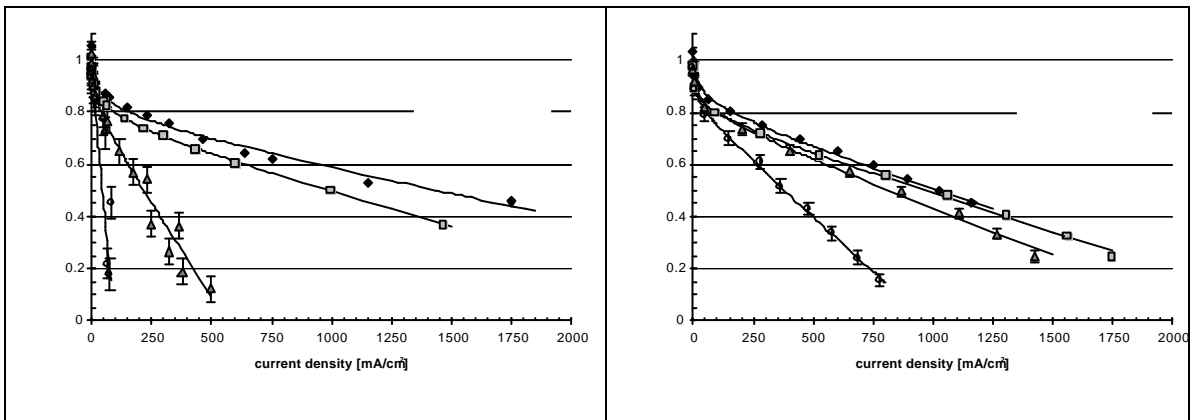

Figure 9A.  PEM Fuel cell performance of an MEA employing a Nafion 115 membrane.  The cell voltage is plotted as a function of average current density in the fuel cell. The cell operating conditions are: ?-1, ?-2, ? -3, ?-4.

Figure 9B.  PEM Fuel cell performance of an MEA employing a Nafion 115/ZP composite membrane.  The cell voltage is plotted as a function of average current density in the fuel cell.  The cell operating conditions are: ?-1, ?-2, ? -3, ?-4.

The MEA polarization curves have been fitted to the equation

$$E = E_{rev} - b\log(\frac{i}{i_o}) - R_{MEA}i \qquad (5)$$

where $E_{rev}$ is the reversible cell potential, b is the Tafel slope, $i_o$ is the exchange current density, i is the current density and $R_{MEA}$ is the membrane electrode resistance. The fit parameters at different operating conditions are summarized in Table 3. The composite membranes show reduced exchange current density compared to the Nafion membranes.

Table 3
Fuel Cell Performance Parameters

| Membrane | Operating Condition | b (mV/dec) | $i_0 \times 10^3$ (mA/cm$^2$) | $R_{MEA}$ (Ω cm$^2$) |
|---|---|---|---|---|
| Nafion 115 | 1 | 84.7 | 3.7 | 0.17 |
| | 2 | 82.0 | 4.6 | 0.24 |
| | 3 | 93.4 | 13.7 | 1.3 |
| | 4 | - | - | 9.8 |
| Nafion 115/ zirconium phosphate (23%) | 1 | 55.0 | 3.0 | 0.3 |
| | 2 | 66.4 | 0.6 | 0.27 |
| | 3 | 59.6 | 0.18 | 0.34 |
| | 4 | 55.1 | 0.06 | 0.79 |

The composite membranes appeared to operate more stably at elevated temperatures. The cell was operated at 130°C for 1h, and then the cell temperature was raised to 140°C. After one hour of stable operation under the latter condition, the cell temperature was returned to 130°C. After this procedure with the composite membrane the current voltage curves returned to their initial (pre-140°C operation) values. The unmodified membranes, by contrast, were altered by the brief exposure (~20 to 30 minutes) to the high temperature. The effective ohmic resistance of the MEA increased from 1.3 O to over 2 O after operation at 140ºC.

## DISCUSSION

When we initiated our studies with composite membranes the hypothesis was that addition of Zirconium phosphate to Nafion would increase the water uptake by the membrane at elevated temperatures. Increasing water content was expected to increase the proton conduction thus improving the membrane performance in the fuel cell. We were surprised that the proton conductivity was diminished in the composite membrane relative to Nafion in spite of the increased water uptake; and we were even more surprised that the fuel cell performance of the composite membranes exceeded Nafion even though the proton conductivity was reduced. Our data does not provide a definite resolution to these conflicting experimental observations, but the data does suggest that mechanical properties of the composite membranes may be the key to the performance of composite membranes.

## *Water Uptake and Ion Exchange Capacity*

The ion exchange capacity of the composite membrane is greater than the ion exchange capacity than Nafion.(Alberti, Casciola et al. 1999; Alberti, Casciola et al. 1999) Zirconium phosphate has exchangable protons, the number of protons depends on the chemical form of the zirconium phosphate. The ion exchange capacity (IEC) for the composite membrane listed in Table 1 can be divided into contributions from the Nafion and from the Zirconium Phosphate.

$$IEC_{composite} = (mass\ fraction\ nafion)IEC_{nafion} + (mass\ fraction\ ZrP)IEC_{ZrP} \qquad (6)$$

Assuming the IEC of nafion in the composite is unchanged the IEC for the Zirconium Phosphate in the membrane is 2670 μeq/g. $Zr(HPO_4)_2$ and $Zr_2(PO_4)_2(HPO_4)$ are well known forms of Zirconium phosphate with ion exchange capacities of 7170 and 2140 μeq/g respectively. The ion exchange capacity of the included zirconium phosphate in the composite membrane is intermediate to these forms of zirconium phosphate; it is easy to imagine a mixture of different zirconium phosphate phases are formed in the composite membrane. XRD identified diffraction peaks that correspond to crystalline zirconium hydrogen phosphate, but most of the zirconium phosphate was amorphous and XRD cannot quantity and forms of the amorphous zirconium phosphate.

Liquid water uptake by Nafion 115 and the Nafion115/ZP composite were the same when normalized by the sulfonic acid concentration. The mass uptake of liquid water normalized by the mass fraction of nafion in the membrane is the same in both membranes.

Water uptake from the vapor is reduced compared to water uptake from liquid water. The composite membranes showed substantially greater water uptake from the vapor phase than water uptake by Nafion 115. The water uptake as a function of water activity, shown in Figure 3, was fit by a finite layer BET isotherm.(Thampan, Malhotra et al. 2000)

$$\lambda = \lambda_m \frac{[ca_w][1 - (n_L + 1)(a_w^{n_l} + n_l a_w^{n_L+1})]}{(1 - a_w)[1 + (c-1)a_w - ca_w^{n_l+1}]} \qquad (7)$$

$\lambda_m$ is the monolayer coverage of water on the sulfonic acid groups, c is related to the chemical potential change due to water adsorption, and $n_L$ is the number of layers that can be adsorbed. The data fits suggest that the more water can be adsorbed in the composite both in the monolayer (first salvation shell –larger $\lambda_m$) and in the multilayer (larger $n_L$); however, the reduction of c for the composite membrane suggests the water is less strongly adsorbed in the composite membrane.

Why should the composite membrane absorb more water from the vapor phase? SAXS data in Figure 6 show the distance between the ionic inclusions in the composite membrane change less than Nafion 115 with increased water content. The scattering peak shifts from 4.5 to 5.1 nm in the composite membrane, and from 3.7 to 5.1 nm with in Nafion 115. The zirconium phosphate is formed when the membrane is fully hydrated, and the distance between the hydrophilic inclusions in at ~5.1 nm. When the membrane is dried, the Nafion contracts but the zirconium phosphate does not. The zirconium phosphate appears to act as scaffold so the membrane cannot shrink much when the water is removed. During rehydration the composite membrane does not swell much because the inorganic scaffold has kept it extended.

The scaffolding effect of the membrane also explains why the water vapor uptake by the composite membrane is greater. When the Nafion 115 membrane absorbs water it must do work to swell the membrane, the composite membrane is already in a partially swelled state, so less work has to be done to swell the membrane, resulting in greater water uptake. The scaffolding effect also explains the reduced density of the dry composite membrane relative to both Nafion and zirconium phosphate. The scaffolding maintains the dimension of the membrane from the swollen state, so as the water is removed void volume is created.

## *Proton Conductivity*

The water content and proton conductivity of Nafion membranes has been studied extensively. The equilibrium water content increases monotonically with increasing water activity of the surrounding phase. Water absorbed in the membrane dissociates the sulfonic acid groups liberating protons that can carry a current. Free protons rarely exist alone in aqueous solution and are almost always bound to one or more waters as hydronium ions; however, we will refer to the mobile ions in the membrane as protons and their motion as proton conductivity. The conductivity depends strongly on the concentration and mobility of the resulting protons. Protons are transported through the membrane by two pathways. The first is a proton shuttling (or Grotthus) mechanism that involves the formation and breaking of hydrogen bonds between the proton and water. Rotation of the water molecules and hydronium ions is required so that the molecules are correctly aligned and new hydrogen bonds can be formed (Bockris and Reddy 1998; Marx, Tuckerman et al. 1999). The second pathway for proton conduction is equivalent to traditional cation conduction where the hydrated proton diffuses through the aqueous media in response to an electrochemical gradient (Zawodzinski, Davey et al. 1995; Eikerling, Kornyshev et al. 1997). The proton conductivity via the Grotthus mechanism is considerably faster than hydronium ion diffusion, and it is estimated that approximately 90% of proton conductance occurs via the Grotthus pathway (Bockris and Reddy 1998).

Proton conductivity depends on the proton concentration, proton mobility, and the conduction path. We fit our proton conductivity data to two models form the literature in an effort to identify the role of the zirconium phosphate in proton conductivity. One model was primarily based on physical properties and a second model based on physical chemistry. The random network model for membrane proton conductivity proposed by Eikerling et al. (Eikerling, Kharkats et al. 1998) is based upon the inverted micelle structure of Nafion and other ionomer membranes. In this model, the inverted micelles, or ionic clusters, are called "pores", and channels connect the random network of these pores distributed throughout the membrane. There are two types of pores described in this model. Wet pores have high water content and consequently they have high conductivity. Dry pores have minimal water, and lower conductivity. As a membrane absorbs water and swells, the fraction of wet pores in the membrane increases, while the fraction of dry pores decreases. With a higher percentage of wet pores the average pore conductivity and the overall bulk membrane conductivity both increase. This model makes use of small angle x-ray scattering (SAXS) data and the cluster swelling model described by Gierke et al. to account for membrane swelling and cluster reorganization with increased water content. (Gierke, Munn et al. 1982; Hsu and Gierke 1982) A random network of wet and dry pores exists in the membrane, whose distributions are governed by the membrane water content and swelling behavior.

The Eikerling model uses a single-bond effective medium approximation (SB-EMA) to solve for the conductivity of this random network. Swelling and structural changes (i.e. ionic cluster reorganization) that occur within the membrane with increasing water content and are described in the following equations:

$$n(w) = n_0(1 + \boldsymbol{a}w) \qquad (8)$$

$$\boldsymbol{u}(w) = \boldsymbol{u}_0(1 + \boldsymbol{b}w)^3 \qquad (9)$$

where $n(w)$ is the number of sulfonic acid groups in an average pore, $n_0$ is the number of sulfonic acid groups in the average pore of a dry membrane, $\boldsymbol{u}(w)$ is the average volume of the pore, $\boldsymbol{u}_0$ is the pore volume in the dry membrane, $w$ is the water content of the membrane in weight percent, and the parameters $\boldsymbol{a}$ and $\boldsymbol{b}$ are used to describe the extent of the swelling and reorganization in the membrane. An increase in parameters $\alpha$ and $\beta$ reflect an increased extent of membrane reorganization. The fraction of wet pores, $x(w)$, depends on the water fraction, the swelling parameters and a scaling parameter, $\boldsymbol{g}$.

$$x(w) = \frac{\boldsymbol{g}w}{(1 + \boldsymbol{b}w)^3 - \boldsymbol{g}w^2\boldsymbol{a}} \qquad (10)$$

The conductivity of the membrane is the weighted average of the conductivities of the wet and dry pores.

$$\boldsymbol{s} = x(w)\boldsymbol{s}_{wet} + (1 - x(w))\boldsymbol{s}_{dry} \qquad (11)$$

The small angle x-ray scattering (SAXS) data and the cluster swelling model proposed by Gierke and Hsu (Gierke and Hsu 1982) were employed to calculate the values for the number of sulfonates per pore and the pore volume ($n$ and $\boldsymbol{u}$) as a function of water content, $w$. The calculated values of $n$ and $\boldsymbol{u}$ were plotted as a function of $w$ and the best fit for parameters $\alpha$ and $\boldsymbol{b}$ were calculated. The parameter $\gamma$ was obtained by equating x(w)=1 when w=0.41 in equation 10. Table 4 shows the parameter values obtained from the SAXS data and conductivity at saturation. A critical caveat in determining these swelling parameters is that the SAXS data was collected at room temperature (25°C), while it is being applied to the membrane conductivity model at 80°C.

Table 4
Proton Conductivity Model of Eikerling

| Membrane | Fitting Parameters | | | | |
|---|---|---|---|---|---|
| | $\alpha$ | $\beta$ | $\gamma$ | $\sigma_{wet}/\sigma_{dry}$ | $\sigma_{x=1}=\sigma_{wet}$ |
| Nafion 115[a] | 0.0693 | 0.034 | 0.10 | 930 | 0.14 |
| Nafion/Zirconium Phosphate[b] | 0.0198 | 0.017 | 0.07 | 600 | 0.12 |

The model fit to the experimental conductivity data is shown in Figure 10. The model consistently overestimates the conductivity. The values of the swelling parameters $\alpha$ and $\beta$ are less for the composite membrane than those obtained for Nafion 115, which reflect the decreased swelling of the composite membrane. The conductivity of the dry pores is nearly an order of magnitude greater for the composite membrane than for Nafion 115. This may be important for fuel cell operation at reduced water activity.

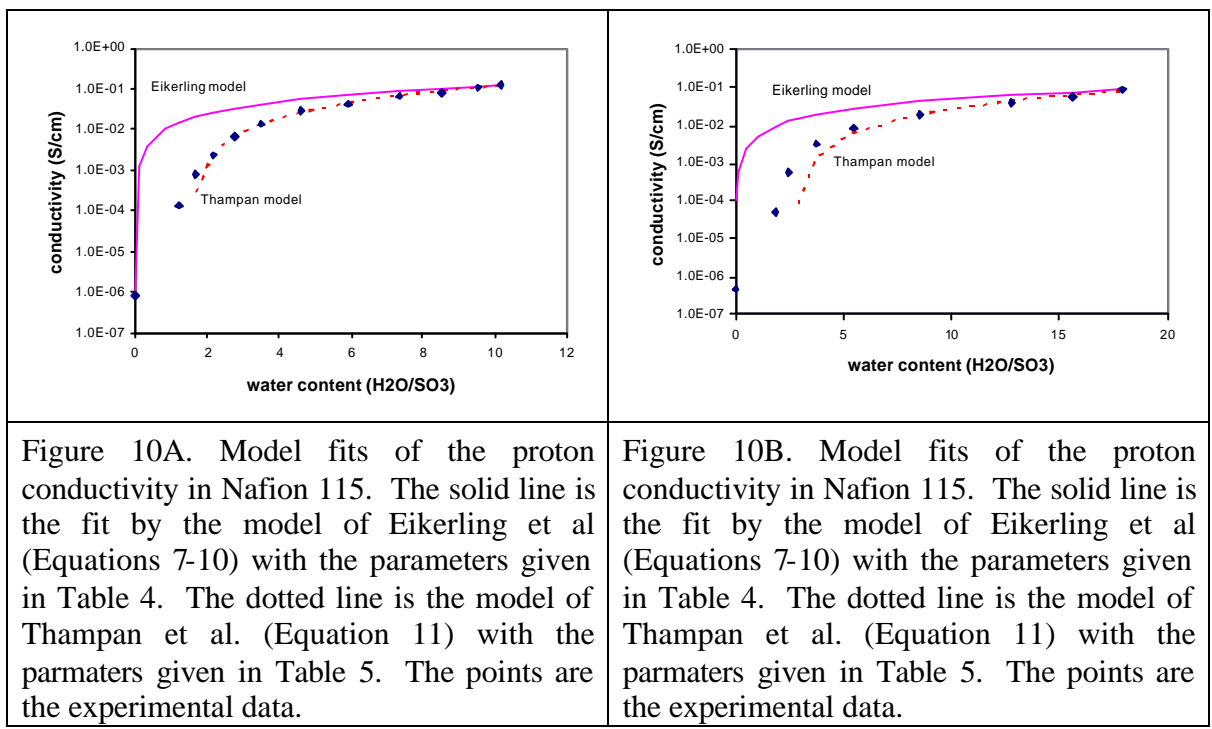

Figure 10A. Model fits of the proton conductivity in Nafion 115. The solid line is the fit by the model of Eikerling et al (Equations 7-10) with the parameters given in Table 4. The dotted line is the model of Thampan et al. (Equation 11) with the parmaters given in Table 5. The points are the experimental data.

Figure 10B. Model fits of the proton conductivity in Nafion 115. The solid line is the fit by the model of Eikerling et al (Equations 7-10) with the parameters given in Table 4. The dotted line is the model of Thampan et al. (Equation 11) with the parmaters given in Table 5. The points are the experimental data.

Thampan et al. presented an alternative model for proton conductivity in ionmers.(Thampan, Malhotra et al. 2000) This model is based upon the similarities between proton diffusion in an aqueous electrolyte and in the membrane electrolyte. Their model introduces corrections to proton diffusion in water to account for the presence of the polymer matrix:

(i)     The additional frictional interaction with the polymer membrane is modeled as proton diffusion through both water and large polymer particles.

(ii)    The introduction of a percolation threshold below which no conduction can take place because of the lack of a continuous conduction pathway.

(iii)   Effective diffusion constants are introduced to account for the porosity and tortuosity of the of polymer matrix.

The model yields the following equations that describe the conductivity of a proton conducting membrane:

$$\boldsymbol{s} = \left(\boldsymbol{e} - \boldsymbol{e}_0\right)^q \left(\frac{\boldsymbol{l}_1^0}{1+\boldsymbol{d}}\right) c_{HA,0} \boldsymbol{a} \qquad (12)$$

where $\varepsilon$ is the fractional membrane volume filled with water and $\varepsilon_0$ is the fractional membrane volume corresponding to the percolation threshold, $q$ is a fitted constant, $c_{HA,0}$ is the concentration of sulfonic acid groups, $\delta$ is the ratio of diffusion coefficients ($D_{H+,H2O}$ and $D_{H+,Membrane}$), and $\lambda_1^0$ is the equivalent conductance at infinite dilution in water. $\alpha$ is the fractional dissociation of the sulfonic acid in the membrane, which is a function of water content.

One of the most crucial elements of this model is the variation of the proton diffusion coefficients with water content. The theoretical relationship between water content and membrane proton diffusion coefficients has not been adequately delineated in the literature, although some attempts have been made on an empirical basis (van der Stegen, van der Veen et al. 1999). Thampan used $\delta$ as a fitting parameter to improve the model fit to literature conductivity data.

The major controls on the membrane conductivity with water content in this model are (i) $\delta$, the ratio of diffusion coefficients, (ii) $c_{HA,0}\alpha$, the proton concentration, (iii) $\varepsilon_0$, the percolation threshold and (iv) q, a fitted parameter which controls the percolation behavior. In our analysis we used the values suggested by Thampan et al. listed in Table 5. The only parameter that we modified between the various membranes is $\delta$. Figures 10A and 10B show the model results compared with experimental conductivity vs. water content data.

Table 5
Proton Conductivity Model of Thampan et al.

| Parameter | Value for Nafion 115 | Value for Nafion/ZP composite |
|---|---|---|
| $\varepsilon_o$ | 0.025 | 0.045 |
| q | 1.5 | 1.5 |
| $c_{HAo}$ | $1.98 \times 10^{-3}/cm^3$ | $2.93 \times 10^{-3}/cm^3$ |
| $K_{dissociation}$ | 6.2 | 6.2 |
| $\lambda_1^o$(at 353 K) | 1650 | 1650 |
| $\delta$ | 0.4 | 3.7 |

The model does a fair job at reproducing the experimental data. As the membrane water content increases the sulfonic acid residues dissociate increasing the concentration of protons in the membrane and increasing the conductivity. The predicted conductivity decreases faster with decreased water content than the experimental data. The composite membranes have significantly increased $\delta$ values compared to unmodified Nafion. The determination of the best fit $\delta$ values suggest that there is a factor of 2-3 decrease in the diffusion coefficient for protons through the "polymer" portion of the composite membrane at given water content. This decrease is much more than the decrease in the water diffusion suggested by the water transport experiments. The water transport through the composite membrane is 35% less than water transport through the Nafion 117 membrane. If water diffusion is limited to the organic phase of the composite the water transport data suggest the zirconium phosphate plays no role in water diffusion. The discrepancy between the water transport data and the model fit for proton conduction suggests that our assumptions of model parameters may not be correct.

Membrane structure is a key membrane characteristic and consequently, a necessary component in a detailed proton conduction model. The model by Thampan et al. could be improved if the relationship of $\delta$ and the membrane structural characteristics were more explicitly determined.

The Thampan model as well as the models of Gieke and Paddison (Paddison 2001) assume a threshold water content below which no proton conduction occurs. Experimentally Nafion membranes have a small but non-zero conductivity even at "zero" water content. The model presented by Thampan et al. neglects tunneling and other diffusion mechanisms that may dominate at low water content.

We have shown a single parameter empirical fit to the conductivity data for Nafion in Figure 4A.

$$\sigma = \sigma_{a_w=0} \exp(c_1 a_w^{c_2}) \tag{13}$$

$\sigma_{aw=0}$ is the conductivity of a dry membrane, $c_1 = \ln(\sigma_{aw=1}/\sigma_{aw=0})$, and $c_2$ is and adjustable parameter. This empirical fit gives a direct evaluation of the proton conductivity based on the thermodynamic water activity of the surrounding vapor phase. This empirical equation fits the conductivity data very well with only one adjustable parameter. The empirical fit is useful for modeling PEM fuel cells where it directly relates the conductivity to the observable variable, water activity.

## Fuel Cell Performance

The most surprising result from our studies is that fuel cell performance, as judged by the polarization curves, is improved for composite membranes even though the conductivity is poorer. This improvement in fuel cell performance appears to be limited to conditions where the water activity is significantly less than one. We estimated the water activity in the fuel cell for different temperatures of the humidifier bottles assuming the same mass transfer efficiency at the different temperatures. The estimated water vapor pressure of the gases humidified at 140º, 130º and 120ºC are 2.4, 2.0 and 1.7 bar respectively, so the water activity in the fuel cell for the three cases are $a_w(140/130/140)=1.0$, $a_w(130/130/130)=0.85$ and $a_w(120/130/120)=0.70$.

Reducing the water activity from 1.0 to 0.7 should increase the membrane resistance for Nafion 115 from 0.20 O-cm$^2$ to 0.22 O-cm$^2$. However, from the fuel cell data the membrane resistance increased from 0.24 O-cm$^2$ to 9.8 O-cm$^2$. The increase in the effective MEA resistance is much greater than predicted from the conductivity measurements. Why is there such a large discrepancy and why does the effective MEA resistance only increase from 0.27 to 0.79 O-cm$^2$ for the composite membrane?

We suggest that the discrepancy is due to the constrained environment of the membrane in the fuel cell. The MEA is compressed between the bipolar plates. The applied pressure on the MEA limits the swelling of the membrane. Absorbed water swells the membrane creating a "swelling pressure" that must overcome the applied sealing pressure of the fuel cell. The greater the water activity the greater the swelling pressure exerted by the membrane. In the fuel cell environment the water content of the membrane is probably much less than expected based on ex-situ measurements because the sealing pressure squeezes water from the membrane.

We suggested the zirconium phosphate in the composite membrane forms a rigid internal scaffolding, which would resist compression of the sealing pressure. At the reduced water activity the composite membrane can take up water and swell without having to overcome the applied sealing pressure. We suggest that the zirconium phosphate provides mechanical strength to the membrane, and it plays little role in the actual conduction of protons through the membrane. This explanation is also consistent with our group's results with other Nafion/metal oxide composite membranes. We have routinely found improved fuel cell performance of the composite membranes at reduced water activity, and there has been little sensitivity to the choice of metal oxide.(Adjemian, Lee et al. 2002; Adjemian, Srinivasan et al. 2002; Costamagna, Yang et al. 2002)

# CONCLUSIONS

Nafion 115/ Zirconium Phosphate composite membranes show enhanced fuel cell performance compared to Nafion 115 at elevated temperature and reduced water activity. A Nafion 115/Zirconium Phosphate composite membrane had a greater ion exchange capacity and took up more water than Nafion 115 membranes, but the composite membranes showed reduced proton conductivity and water transport. Small angle X-ray scattering data indicated the spacing between hydrophilic phases in the composite membrane were further apart than in Nafion 115, and there appeared to be less restructuring of the composite membrane with water absorption. The data suggest that the Zirconium Phosphate forms an internal rigid scaffold within the membrane that permits increased water uptake by the membrane in the confined environment of the fuel cell Membrane Electrode Assembly.